\documentclass[]{aa}
\usepackage{txfonts}
\usepackage{graphicx}
\newcommand{\R}{{\mathbb R}}
\begin{document}

\title{On the origin of $rR_1$ ring structures in
barred galaxies}
\author{M. Romero-G\'omez\inst{1,3}, J.J. Masdemont\inst{2},
E. Athanassoula\inst{3}, \and C. Garc\'{\i}a-G\'omez\inst{1}}
\institute{D.E.I.M., Universitat Rovira i Virgili, Campus Sescelades, Avd. dels Pa\"{\i}sos
Catalans 26, 43007 Tarragona, Spain
\and I.E.E.C \& Dep. Mat. Aplicada I, Universitat Polit\`ecnica de
Catalunya, Diagonal 647, 08028 Barcelona, Spain
\and LAM, Observatoire Astronomique de Marseille Provence, 
2 Place Le Verrier, 13248 Marseille C\'edex 04, France}
\offprints{M. Romero-G\'omez, \email{merce.romero@urv.net}}
\date{Received 15 August}
\abstract{We propose a new theory for the formation of $rR_1$ ring
structures, i.e. for ring structures with both an inner
  and an outer ring, the latter having the form of
  ``8''. We propose that these rings are formed by material from the
stable and unstable invariant manifolds associated with the Lyapunov
orbits around the equilibrium points of a barred galaxy. We discuss
the shape and velocity structure of the rings thus formed and argue
that they are in agreement with the observed properties of $rR_1$ structures.
}
\authorrunning{M. Romero-G\'omez et al.}
\titlerunning{Origin of $rR_1$ galaxies}
\maketitle

\section{Introduction}

\begin{figure}
\sidecaption
\includegraphics[scale=0.5,angle=0.0]{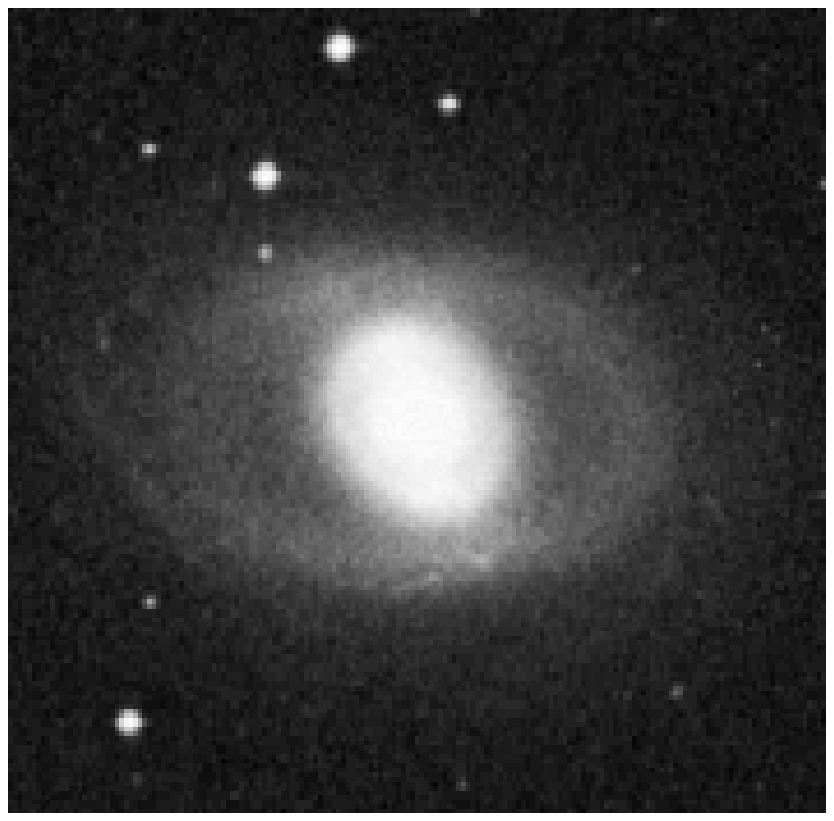}
\caption{ Image of the $rR_1$ galaxy NGC 1326 showing a
well developed $(R_1)SB(r)0/a$ structure. (Digital Sky Survey \copyright Anglo
Australian Observatory Board)}
\label{fig:ngc1326}
\end{figure}

Barred galaxies often show spectacular rings whose different types
have been classified by Buta (\cite{but86}) as follows:
Nuclear rings (which are not discussed
here) which surround the nucleus and that are much smaller in size than the
bar. At larger radii there are inner rings, denoted in Buta's
(\cite{but86}) classification by a ``$r$'', which surround the bar
and which have the same size and orientation as the bar. And there are
outer rings, denoted by $R$, which are bigger than the bar. These
``pure'' rings are defined to be distinct and closed, but one can often
find unclosed or partial ring patterns of spiral character, and these
are referred to as ``pseudorings'' and denoted $R'$.
A particular class of outer rings called $R_1$ or ${R_1}'$, for
pseudorings, has two main arms forming an
eight-shaped ring or pseudoring, with its major axis perpendicular to
the bar. NGC 1326 is a well studied example of an $(R_1)SB(r)0/a$ galaxy
(Buta \cite{but95}). Its bar is
surrounded by an inner ring which is
almost exactly aligned with the bar and has roughly the same
diameter. Its outer ring is clearly $R_1$ and is elongated perpendicular to the bar
(Buta et al. \cite{but98}, see Fig. \ref{fig:ngc1326}). 

Schwarz (\cite{schw81}, \cite{schw84}, \cite{schw85}) showed that
ring-like structures can arise around the Lindblad resonances due to
a bar-like perturbation of the galaxy potential. The gas will be
forced to rearrange its distribution and generate a spiral. Near the
outer Lindblad resonances the crossing of perturbed trajectories
will develop a ring-like pattern. In these regions, gas
clouds will collide and will form spiral shock fronts which will 
slowly change as a result of torques exerted by the bar and
evolve into a ring structure which, after star formation, will be 
populated by stars in near-resonant periodic orbits. 

Danby (\cite{dan65}) argued that orbits in the gravitational potential
of a bar play an important role in the formation of arms. He noted that orbits departing from the
vicinity of the equilibrium points located at the ends of the bar
describe loci with the shape of
spiral arms and can be responsible for transport of stars from within
to outside corotation, and viceversa. Unfortunately, he did not set
his work in a rigorous theoretical context, so that it remained purely
phenomenological. He also investigated whether orbits can be
responsible for ring-like structures, but, in this case, he did not consider orbits
departing from the ends of the bar as he previously did when
accounting for the spiral arms. He concluded that rings would require
high energy orbits, while we will see in this
paper that mainly low energy orbits can constitute the rings.

Here we propose a new dynamical model, applicable to the
particular case of $rR_1$ class of ringed galaxies. We expect
that more detailed modelling will extend this new model
to the rest of the ringed galaxy classes. The model is based
on the orbital motion of stars in the vicinity of equilibrium points
in the rotating bar potential and does not rely on additional star
formation. The tools we use  in our model are well known in celestial
mechanics but have not been much
used so far in galactic dynamics. We therefore start, in section 2, by
introducing the equations of motion and the equilibrium
points. In section 3 we describe the dynamics around the Lagrangian
point $L_1$ and we introduce the invariant manifolds. We discuss the
linear case in subsection 3.1 and the general case in subsection 3.2. In section 4 we
describe the role of the invariant manifolds in the transport of stars
and their properties in the framework of ringed barred galaxies.
In section 5 we describe briefly how invariant manifolds are computed
numerically and we apply them to a specific model. Finally, section
6 discusses the properties of the spiral arms thus produced and
summarises our results.

\section{Equations of motion and equilibrium points}

We model the potential of the galaxy as the superposition of two
components, one axisymmetric and the other barlike.
The latter rotates clockwise at angular velocity
${\bf\Omega_p}=\Omega_p{\bf e_z}$, where $\Omega_p>0$ is the
pattern speed considered here to be constant\,\footnote{Bold letters
denote vector notation.}. 
The equations of motion in the frame rotating with ${\bf \Omega_p}$ are:
\begin{equation}\label{eq-motvec}
{\bf
\ddot{r}=-\nabla \Phi} -2{\bf (\Omega_p \times \dot{r})-  \Omega_p \times
(\Omega_p\times r)},
\end{equation}
where the terms $-2 {\bf \Omega_p\times \dot{r}}$ and $-{\bf \Omega_p \times
(\Omega_p\times r)}$ represent the Coriolis and the centrifugal
forces, respectively, and ${\bf r}$ is the position vector. 

Following Binney \& Tremaine (\cite{bint87}), we take
the dot product of equation (\ref{eq-motvec}) with
${\bf \dot{r}}$, and rearranging the resulting equation, we obtain
$$
\frac{d E_J}{dt}=0,
$$
where 
$$
E_J \equiv \frac{1}{2} {\bf\mid\dot{r}\mid}^2 + \Phi -\frac{1}{2}\mid {\bf\Omega_p
\times r}\mid^2.
$$
$E_J$ is known as the Jacobi integral or Jacobi constant. Notice that
this is the sum of $\frac{1}{2} {\bf\dot{r}^2} + \Phi$, which is
the energy in a nonrotating frame, and of the quantity \\
\noindent \hbox{$ -\frac{1}{2}\mid {\bf\Omega_p\times
r}\mid^2=-\frac{1}{2}\Omega_p^2\,(x^2+y^2)$}, 
which can be
thought of as the ``potential energy'' to which the centrifugal
``force'' gives rise. So if we define an effective potential
$$\Phi_{\hbox{\scriptsize eff}}=\Phi-\frac{1}{2}\Omega_p^2\,(x^2+y^2),
$$ equation (\ref{eq-motvec}) becomes
\begin{equation}\label{eq-motvec2}
{\bf \ddot{r}=-\nabla \Phi_{\hbox{\scriptsize eff}}} -2{\bf (\Omega_p \times \dot{r})},
\end{equation}
and the Jacobi constant is 
$$
E_J = \frac{1}{2} {\bf\mid \dot{r}\mid} ^2 + \Phi_{\hbox{\scriptsize eff}},
$$
so that it can be considered as the ``energy'' in the
rotating frame.

\begin{figure*}
\begin{center}
\includegraphics[scale=0.8,angle=-90.0]{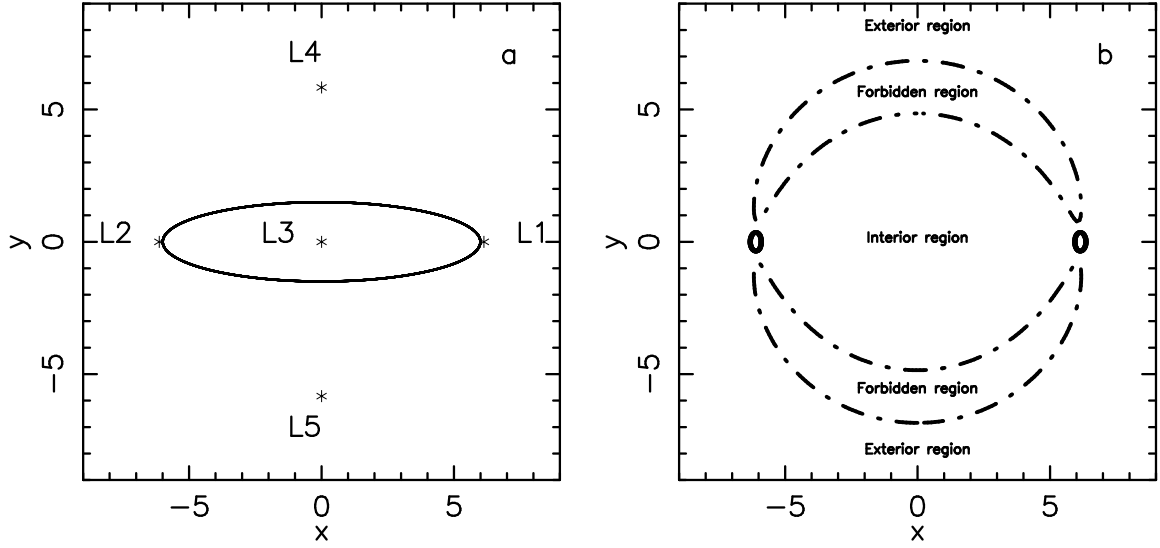}
\end{center}
\caption{{\bf a)} Outline of the bar and the position of the five equilibrium points
(marked with a star). {\bf b)} The two Lyapunov orbits (thin solid lines) and the zero
velocity curves (dot-dashed lines) delimiting the forbidden, the interior and
exterior regions. Both are given for the same energy value.}
\label{fig:system}
\end{figure*}

The surface $\Phi_{\hbox{\scriptsize eff}}=E_J$ is called the zero
velocity surface and its cut with the $z=0$ plane is the zero velocity
curve. All regions in which $\Phi_{\hbox{\scriptsize
eff}}>E_J$ are forbidden to a star and we call them forbidden
regions. Fig. \ref{fig:system}b shows an example of zero
velocity curves and the regions delimited by them, namely the exterior,
interior and forbidden regions, for the potential introduced in
section 5.

$\Phi_{\hbox{\scriptsize eff}}$ has five
equilibrium points, named $L_1$ to $L_5$, located in the $xy$
plane, at which 
$\frac{\partial \Phi_{\hbox{\scriptsize eff}}}{\partial x}=
\frac{\partial \Phi_{\hbox{\scriptsize eff}}}{\partial y}=
\frac{\partial \Phi_{\hbox{\scriptsize eff}}}{\partial z}= 0$. 
Due to their similarity to the corresponding points in the 
restricted three body problem, they are often called Lagrangian
points. $L_1$ and $L_2$ lie on the $x$-axis and are symmetric with
respect to the origin. $L_3$ lies on the origin of
coordinates. Finally, $L_4$ and $L_5$ lie on the $y$-axis and
are also symmetric with respect to the origin (see
Fig.\ref{fig:system}a, again for the potential introduced in section 5).

We can check the stability of the Lagrangian points by considering the
motion in their immediate neighbourhood\,\footnote{We can refer to Pfenniger
(\cite{pfe90}), where he studied the stability character of the
Lagrangian points in different stellar bars.}. If we expand
$\Phi_{\hbox{\scriptsize eff}}$ 
around one of these points and retain only first order terms, the equations of motion
(\ref{eq-motvec2}) become 
\begin{equation}\label{eq-mot}
\left\{
\begin{array}{rcl}
\ddot{x}  & = & 2\Omega_p\dot{y} -\Phi_{xx}x \\
\ddot{y}  & = & -2\Omega_p\dot{x}-\Phi_{yy}y\\
\ddot{z} & = & -\Phi_{zz}z
\end{array}
\right.
\end{equation}
where we have defined $$ x \equiv x-x_{\hbox{\scriptsize L}};\quad y \equiv y-y_{\hbox{\scriptsize L}};\quad z \equiv z-z_{\hbox{\scriptsize L}}, $$
$$\Phi_{xx}\equiv
\left ( \frac{\partial^2\Phi_{\hbox{\scriptsize eff}}}{\partial
x^2}\right )_{L_i} ; \quad \Phi_{yy}\equiv
\left ( \frac{\partial^2\Phi_{\hbox{\scriptsize eff}}}{\partial
y^2}\right )_{L_i}; \quad \Phi_{zz}\equiv
\left ( \frac{\partial^2\Phi_{\hbox{\scriptsize eff}}}{\partial
z^2}\right )_{L_i} $$
and $x_{\hbox{\scriptsize L}}$, $y_{\hbox{\scriptsize L}}$ and
$z_{\hbox{\scriptsize L}}$ are the coordinates of the Lagrangian point
$L_i$.
Note that, for any barlike potential whose principal axes lie along the
coordinate axes, $\left ( \frac{\partial^2\Phi_{\hbox{\scriptsize eff}}}{\partial
x\partial y}\right )_{L_i}=0$ by symmetry.
Setting $x_1=x,\quad x_2=y,\quad x_3=z,\quad x_4=\dot{x},\quad
x_5=\dot{y}$ and $x_6=\dot{z}$, equations (\ref{eq-mot}) are written
as a system of first order differential equations,
\begin{equation}\label{eq-first}
\left \{ \begin{array}{rclcl}
\dot{x_1} & = & f_1(x_1,\dots,x_6) & = & x_4\\
\dot{x_2} & = & f_2(x_1,\dots,x_6) & = & x_5\\
\dot{x_3} & = & f_3(x_1,\dots,x_6) & = & x_6\\
\dot{x_4} & = & f_4(x_1,\dots,x_6) & = & 2\Omega_p\, x_5-\Phi_{xx}x_1\\
\dot{x_5} & = & f_5(x_1,\dots,x_6) & = & -2\Omega_p\, x_4-\Phi_{yy}x_2\\
\dot{x_6} & = & f_6(x_1,\dots,x_6) & = & -\Phi_{zz}x_3.
\end{array} \right.
\end{equation}

\section{Dynamics around $L_1$. The invariant manifolds}
\subsection{Linear case}

Let us now focus on the dynamics around $L_1$ ($L_2$
is completely symmetrical). The differential matrix associated to
system (\ref{eq-first}) around $L_1$ is
$$ 
Df_x(L_1)=\left ( \begin{array}{cccccc}
0 & 0 & 0 & 1 & 0 & 0\\
0 & 0 & 0 & 0 & 1 & 0\\
0 & 0 & 0 & 0 & 0 & 1\\
-\Phi_{xx} & 0 & 0 & 0 & 2\Omega_p & 0\\
0 & -\Phi_{yy} & 0 & -2\Omega_p & 0 & 0\\
0 & 0 & -\Phi_{zz} & 0 & 0 & 0
\end{array} \right )
$$
We obtain the stability character of $L_1$ by studying the eigenvalues
of this matrix. It has six eigenvalues: $\lambda$, 
$-\lambda$, $\omega i$, $-\omega i$,  $\nu i$ and $-\nu i$, where $\lambda$,
$\omega$ and $\nu$ are positive real numbers, i.e. $L_1$ is a 
linearly unstable point. The corresponding eigenvectors (either real or
complex) have zero $x_3$, and $x_6$ components in the case of $\pm \lambda$ and
$\pm \omega i$, and zero $x_1,x_2,x_4$, and $x_5$ components in the case of
$\pm \nu i$. Since the purely imaginary eigenvalues
denote oscillation and the real eigenvalues are associated to
a saddle behaviour, i.e. exponential behaviour with opposite exponents, the
linearised flow around $L_1$ in the 
rotating frame of coordinates is characterised by a superposition of an 
harmonic motion in the $xy$ plane (equatorial plane), a saddle
behaviour in this plane, and an oscillation in the $z$-direction.

Because of this unstable character, the equilibrium points $L_1$ and 
$L_2$ set the limits of the stability region around the
stable point $L_3$, and thus, they set an upper limit to the extension
of the bar. This, however, is only an upper limit and in most
cases the bar is much shorter than that limit (Athanassoula \cite{ath92};
Patsis, Skokos \& Athanassoula \cite{pat03}). 

The central stable point, $L_3$, is surrounded by the
classic $x_1$ family of periodic orbits which is responsible for maintaining the bar
structure, while the stable points $L_4$ and $L_5$ are surrounded by 
families of periodic banana orbits (Contopoulos \& Papayannopoulos
\cite{CP}; Athanassoula et al. \cite{atha83}; Contopoulos
\cite{con81}; Skokos, Patsis \& Athanassoula
\cite{sko02}). All these orbits have been well studied for many 
models (Contopoulos \cite{con02} and references therein)  and we will
not discuss them here any further. 

As already mentioned, the general linear motion  
around $L_1$ is obtained by the addition of an hyperbolic exponential 
part to the in-plane and out-of-plane oscillations mentioned. We note
that this exponential part has both stable and unstable components 
with exponents of opposite sign. Following again Binney \& Tremaine (\cite{bint87}) we write
\begin{equation}\label{eq:sol-lingen}
\left\{
\begin{array}{l}
x(t)  =  X_1 e^{\lambda t} + X_2 e^{-\lambda t}  
        +X_3\cos(\omega t + \phi), \\
y(t)  = X_4 e^{\lambda t} + X_5 e^{-\lambda t}
        +X_6\sin(\omega t +\phi),\\
z(t)  =  X_7\cos(\nu t +\psi).
\end{array}
\right.
\end{equation}
Here $X_i, i=1,\dots, 7$ and $\phi$, $\psi$ are values representing
amplitudes and phases. Substituting these equations into the
differential equations (\ref{eq-mot}), we find that $X_i$ are related by
\begin{equation}
\begin{array}{rcccl}
X_4 & = & \frac{\Phi_{xx}+\lambda^2}{2\Omega_p\lambda} X_1 & = &
-\frac{2\Omega_p\lambda}{\Phi_{yy}+\lambda^2}X_1 \rule[-.5cm]{0cm}{1cm}\\
X_5 & = & -\frac{\Phi_{xx}+\lambda^2}{2\Omega_p\lambda}X_2 & = &
\frac{2\Omega_p\lambda}{\Phi_{yy}+\lambda^2}X_2\rule[-.5cm]{0cm}{1cm}\\
X_6 & = & \frac{\Phi_{xx}-\omega^2}{2\Omega_p\omega}X_3 & = &
\frac{2\Omega_p\omega}{\Phi_{yy}-\omega^2}X_3\rule[-.5cm]{0cm}{1cm}
\end{array}
\end{equation}
We define $A_1=\frac{\Phi_{xx}+\lambda^2}{2\Omega_p\lambda}$ and
 $A_2=\frac{\Phi_{xx}-\omega^2}{2\Omega_p\omega}$. Note that $A_1$
depends only on $\lambda$ and $A_2$ depends only on
$\omega$. Moreover  $X_4=A_1 X_1$,
$X_5=-A_1X_2$, and $X_6=A_2X_3$, so that equation (\ref{eq:sol-lingen}) becomes
\begin{equation}\label{eq:sol-lin}
\left\{
\begin{array}{l}
x(t)  =  X_1 e^{\lambda t} + X_2 e^{-\lambda t}  
        +X_3\cos(\omega t + \phi), \\
y(t)  =  A_1X_1 e^{\lambda t} - A_1X_2 e^{-\lambda t}
        +A_2X_3\sin(\omega t +\phi),\\
z(t)  =  X_7\cos(\nu t +\psi).
\end{array}
\right.
\end{equation}

In the sequel we will restrict ourselves to the motion in the
equatorial plane (i.e. $z=0$, or $X_7=0$). 
This restriction is not critical in the dynamics we want to study, since 
the $z$ component essentially only adds a vertical oscillation to 
the planar motion. For a study of the vertical orbital structure
around the Lagrangian points in barred galaxies, see Oll\'e \&
Pfenniger (\cite{oll98}).

Using (\ref{eq:sol-lin}), a given state $(x,y,\dot{x},\dot{y})$ at $t=0$ 
is characterised by a choice of $(X_1,X_2,X_3,\phi_1)$, modulus $2\pi$
in the phase $\phi$. When $X_1=X_2=0$ the
initial condition is of the form 
\begin{eqnarray*}
(x(0),y(0),\dot{x}(0),\dot{y}(0))=~~~~~~~~~~~~~~~~~~~~~~~~~~~~~~~~~~~~~ \\
~~(X_3 \cos\phi, A_2X_3 \sin\phi, -X_3\omega \sin\phi,
A_2X_3 \omega\cos\phi) 
\end{eqnarray*}
for selected values of $X_3$ and $\phi$. 
When time evolves, we obtain from this initial condition the periodic
motion,
$$
\begin{array}{ll}
{\bf x}_0(t)=& (x,y,\dot{x},\dot{y})=\\
&(X_3 \cos(\omega t+\phi),\, A_2X_3 \sin(\omega t+\phi), \\
&\,-X_3\omega\sin(\omega t+\phi),\, A_2X_3\omega\cos(\omega t+\phi)),
\end{array}
$$

\noindent of period $\tau$ to which we will refer as a linear Lyapunov periodic orbit.

Consider now any small deviation ${\bf \delta}$ from the periodic orbit
${\bf x}_0(t)$, ${\bf x}(t)={\bf x}_0(t)+{\bf \delta}$. Inserting this
into the equations of motion (\ref{eq-first}), and
linearising them with respect to ${\bf \delta}$, we obtain the
variational equations
\begin{equation}\label{vareq}
\dot{\delta}=\frac{\partial {\bf f}}{\partial {\bf
x}}\delta=\sum_{i=1}^{6}\left ( \frac{\partial f_k}{\partial
x_i}\right)_{{\bf x}_0}\delta_i
\end{equation}
as defined in Contopoulos (\cite{con02}), where 
$$
A(t)=\left( \delta_{ik} \right) = \left(\frac{\partial f_k}{\partial
x_i}\right)=\left( \begin{array}{ccc}
\frac{\partial f_1}{\partial x_1} & \ldots & 
\frac{\partial f_1}{\partial x_6}\\
\vdots & \ddots & \vdots\\
\frac{\partial f_6}{\partial x_1}  &\ldots & 
\frac{\partial f_6}{\partial x_6}
\end{array}\right)
$$
is the variational matrix. The variational equations are linear
equations in $\delta_i\, (i=1,\dots,6)$ with periodic coefficients of
period $\tau$. These equations are used to study the stability character of
a periodic orbit. If we integrate the variational equations
(\ref{vareq}) until time $\tau$ with initial conditions
$\delta_{1k}=(1,0,\dots,0),\dots,\delta_{6k}=(0,\dots,0,1)$, we obtain
the monodromy matrix. The eigenvalues and eigenvectors of the
monodromy matrix give us information on the stability character of a
periodic orbit.

Returning to equations (\ref{eq:sol-lin}), let us consider a
similar initial condition but with $X_1= 0$ and
$X_2\neq 0$. According to (\ref{eq:sol-lin}) the exponential term 
proportional to $X_2$ vanishes when time tends to infinity and the trajectory
tends to the linear Lyapunov. All these type of orbits form what is called 
the stable manifold of the linear Lyapunov. In the same way, if the initial 
condition is chosen with $X_1 \neq 0$, $X_2=0$ the 
exponential term proportional to
$X_1$ tends to zero when time tends to minus infinity, and all
such orbits form
what is called the unstable manifold of the linear Lyapunov.
Roughly speaking, orbits in the stable/unstable manifold are asymptotic
orbits, which tend to/depart from the linear Lyapunov orbit.

\subsection{General case}

All the definitions given in section 3.1 for the linear case are easy
to extend to the general case when the full equations of
motion are considered. From $L_1$ emanates a 
family of planar periodic
orbits known as Lyapunov orbits (Lyapunov \cite{lya49}) which locally can
be parametrised by the energy. The full set of Lyapunov orbits form a
family of periodic orbits, which we will denote by $\Gamma$.

\begin{figure}
\begin{center}
\includegraphics[scale=0.5,angle=0.0]{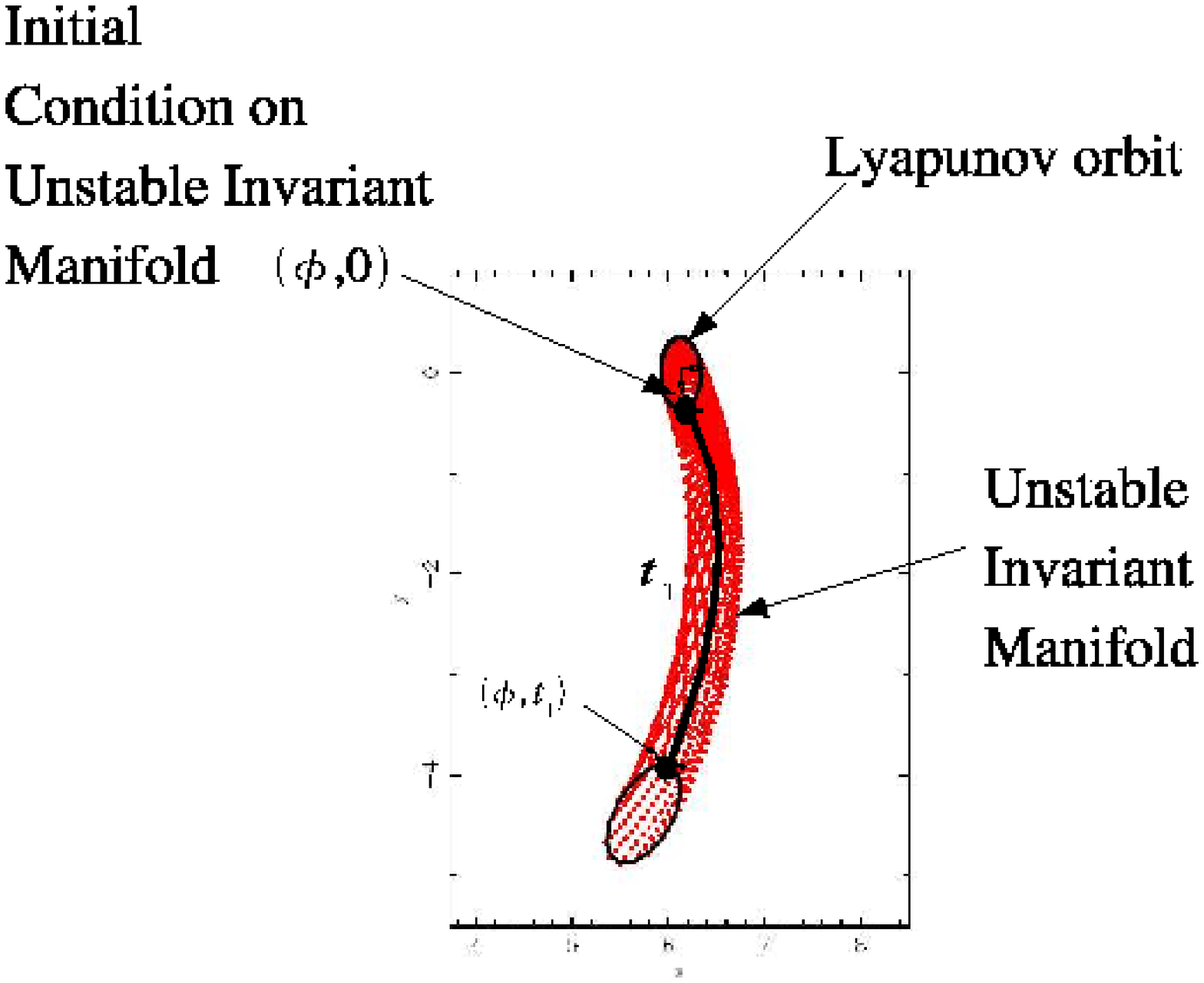}
\end{center}
\caption{Schematic view of an unstable invariant manifold.}
\label{fig:cilinder}
\end{figure}

\begin{figure}
\begin{center}
\includegraphics[scale=0.35,angle=-90.0]{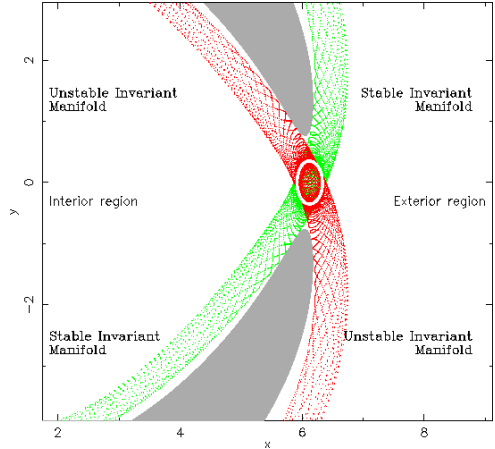}
\end{center}
\caption{Invariant manifolds. In the centre of the plot, and in white solid line, the
Lyapunov orbit around $L_1$. The two
branches of the unstable invariant manifold (red or dotted lines), and the two branches of the stable
invariant manifold (green or dotted lines). In grey, the forbidden region surrounded by the zero velocity curves.}
\label{fig:branches}
\end{figure}

Let us denote by $\Psi(t,X)$ the orbit that at $t=0$ has the state
$X=(x,y,\dot{x},\dot{y})$. For a given $\gamma \in \Gamma$, i.e. for 
a given Lyapunov orbit, we define the stable manifold of $\gamma$ as
$$
W^s_\gamma=\left\{ X\in\R^4 \quad / 
    \lim_{t\rightarrow\infty}||\Psi(t,X)-\gamma||=0\right\},
$$
where the double bars denote Euclidean distance. Thus, simply
speaking, the stable manifold is the set of orbits which
tend to the Lyapunov orbit as time tends to infinity. In the same way, the 
unstable manifold of $\gamma$ is defined as
$$
W^u_\gamma=\left\{ X\in\R^4 \quad / 
    \lim_{t\rightarrow -\infty}||\Psi(t,X)-\gamma||=0\right\}.
$$
Simply speaking again, the unstable manifold is the set of orbits
which tend to the Lyapunov orbit as time tends to minus infinity, or, equivalently,
the set of orbits departing from the Lyapunov.
The orbits of $W^s_\gamma$ and $W^u_\gamma$ have the same energy as 
the Lyapunov orbit $\gamma$, and so they belong to the same energetic 
three dimensional manifold where 
$\gamma$ is contained. Moreover, $W^s_\gamma$ and $W^u_\gamma$ are 
two dimensional tubes, which, similarly to the former linear 
example, can be parametrised by the angle $\phi$ and the time
$t$ (see Fig.~\ref{fig:cilinder} and Masdemont (\cite{masde05})). 
We also note that both $W^s_\gamma$ and $W^u_\gamma$ have two branches 
meeting at the Lyapunov orbit in a way similar to a saddle point (see
Fig.~\ref{fig:branches}).

\section{The role of invariant manifolds}

\begin{figure}
\begin{center}
\includegraphics[scale=0.3,angle=-90.0]{curvebn.ps}
\includegraphics[scale=0.5,angle=0.0]{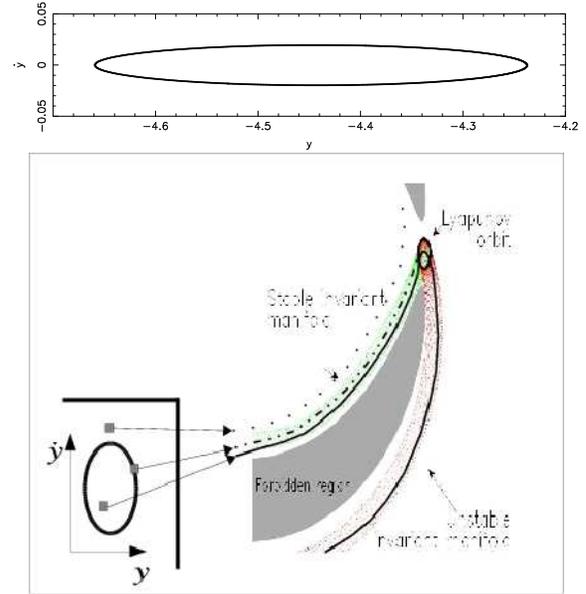}
\end{center}
\caption{Transport of material. {\bf Top panel:} the curve $W^s_{\gamma,1}$ in the $y\dot{y}$
plane. {\bf Bottom panel:} schematic view of the dynamics in the region around the
$L_1$ Lagrangian point. In grey, 
the region delimited by the zero velocity
curves. The dotted line gives a non-transit orbit, which is confined to the interior
region and whose intersection with the $y\dot{y}$ plane is
located outside the $W^s_{\gamma,1}$ curve; the solid line, a transit
orbit, which in the $y\dot{y}$ plane is
located inside the $W^s_{\gamma,1}$ curve; and the dot-dot-dash line, an
asymptotic orbit of the stable invariant manifold. In the
inlay, a schematic view of the curve $W^s_{\gamma,1}$ in the $y\dot{y}$
plane with the location of the three orbits.}
\label{fig:traj}
\end{figure}

Invariant manifolds of Lyapunov orbits play a crucial role in the transport
of material between different parts of the configuration
space. Lyapunov orbits
are located near the ends of the
bar of the galaxy, between the two banana like zero velocity curves which 
surround the forbidden regions of motion for the considered energy,
and, loosely speaking, can be considered as gates between
the interior and exterior region they delimit (see Fig \ref{fig:system}b).
Let us consider for instance the stable manifold, $W^s_\gamma$, inside
the interior region and integrated backwards in time till it
crosses the plane, $S$, defined by $x=0$. The plot of this 
intersection in the $y\dot{y}$ plane is a closed curve
$W^s_{\gamma,1}$ (Fig. \ref{fig:traj}).
Each point in the plane $y\dot{y}$ of $S$ 
corresponds to a given trajectory, since $x=0$ by
the definition of $S$ and $\dot{x}$ can be obtained from the 
condition that the energy of the state $(x,y,\dot{x},\dot{y})$ 
be the selected one (the sign of $\dot{x}$ is determined by the
sense of crossing). $W^s_{\gamma,1}$ is a closed curve that splits the $y\dot{y}$ plane 
in $S$ in three different regions:
the curve itself, the points exterior to the curve and the points
interior to the curve. By definition, the points on the
curve $W^s_{\gamma,1}$ belong to $W^s_{\gamma}$ and are therefore orbits that
tend asymptotically to the Lyapunov orbit. The points outside $W^s_{\gamma,1}$
are states whose trajectories remain inside the interior region of the
galaxy delimited by the zero velocity curves, while the 
points inside $W^s_{\gamma,1}$ correspond to orbits that transit from the 
interior region to the exterior one. These last orbits, the transit orbits,
are confined inside
the tube $W^u_\gamma$ and, as we will argue in the following sections,
are the orbits which form part of the rings of
the galaxy for the considered energy value. In this way, the
manifolds of the Lyapunov orbits drive the motion of the stars from 
the interior to the exterior regions. For more details of this mechanism,
although in another context, see G\'omez et al. (\cite{gom04}) and references
therein. Since these invariant manifolds are not limited to the vicinity 
of the unstable points, but extend well beyond it, they can be responsible
for global structures and we will argue in this paper that they, together with
the orbits driven by them, could be responsible for the 
ring structures in barred galaxies.

\section{Application to a ringed barred galaxy model}

In this section we will explicitly calculate the invariant manifolds
in a barred galaxy model. The reader who has only skimmed the previous
sections should keep in mind that the invariant manifolds are just ensembles of
orbits linked to the $L_1$ and $L_2$ Lyapunov orbits.

 Since invariant manifolds can only be calculated 
numerically, we first adopt a simple, yet
realistic, barred galaxy model (Pfenniger \cite{pfe84}). 
Our model bar consists of an axisymmetric component, modelled by a
Miyamoto-Nagai potential (Miyamoto \& Nagai
\cite{miya75})
$$
\Phi_d=-\frac{GM_d}{\sqrt{x^2+y^2+(A+\sqrt{B^2+z^2})^2}},\rule[-.5cm]{0cm}{1cm}
$$
and a Ferrers bar (Ferrers \cite{fer77}) 
$$
\rho=\left \{\begin{array}{lr}
\rho_c(1-m^2)^n & m\le 1\\
 0 & m\ge 1,
\end{array}\right.
$$
where $m^2=x^2/a^2+y^2/b^2+z^2/c^2$ and
$\rho_c=\frac{105}{32\pi}\frac{GM_b}{abc}$ is the central
density. We
take $A= 3$, $B = 1$, $n = 2$, $a= 6$, 
$b = 1.5$, $c = 0.6$, $GM_d = 0.9$ and $GM_b = 0.1$. 
The pattern speed
is taken such as to place corotation at the end of the bar. The length
unit is the $kpc$, the total mass $G(M_d+M_b)$ is set to be equal to
1, and the time unit is $2\times 10^6\,yr$. With these units, the value of
the potential energy at $L_3$, $L_1 (L_2)$, and $L_4 (L_5)$ is
-0.31503, -0.19789 and -0.19456, respectively. As done already in the previous sections, we
limit ourselves to the $z$ = 0 plane, since the instability we are
interested in is contained to this plane and the $z$ component
only adds vertical oscillations, which are unimportant in this context. 

In this model, we computed the invariant manifolds numerically using
an approach similar to the
one of G\'omez et al. (\cite{gom93}), i.e. we use a linear approach 
to obtain the initial conditions of the orbits that constitute the
invariant manifolds\,\footnote{Higher order approximations could be
obtained using similar techniques as in Masdemont (\cite{masde05}),
but with much more difficulty and regularity problems.}. As previously
mentioned, the linear motion around $L_1$ and $L_2$ consists of an
hyperbolic exponential part in the in-plane and of an out-of-plane
oscillation. The exponential part has both stable and unstable
components, which correspond to the stable and unstable eigenvectors
of the monodromy matrix of the Lyapunov orbits around $L_1$ and
$L_2$. Therefore, we obtain the initial conditions for the stable and
unstable invariant manifolds shifting positions and velocities of the
Lyapunov orbit by a small amount ($10^{-5}$) in the direction given by the stable
and unstable eigenvectors, respectively.
The global extension of the manifold is then obtained by integrating
numerically with a Runge-Kutta-Felhberg of orders 7-8.
\begin{figure}
\begin{center}
\includegraphics[scale=0.45,angle=-90]{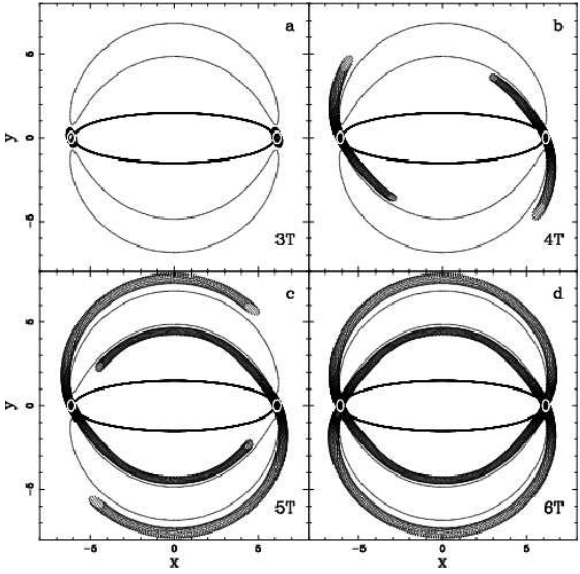}
\end{center}
\caption{Unstable invariant manifolds for four
  different times given in the lower right corner of each panel. T is
the period of the bar rotation.}
\label{fig:times}
\end{figure}

Figure \ref{fig:times} shows the evolution of the length of the
unstable invariant manifolds
with time. Here, $T$ is the rotation period of the bar. Since the
invariant manifolds consist of a set of orbits, their time evolution
is that of an orbit. So Fig. \ref{fig:times} shows the
orbits composing the invariant manifolds integrated up to 3,4,5 and 6 bar
rotations, respectively, in order to check their evolution with time. Note that they
leave the vicinity of the Lyapunov orbits, and stay close to the zero
velocity curves. When they get to the $x=0$ axis, they bend and head towards
the opposite side of the bar, completing the
ring structure. Thus our calculations show that the apocenter of the
orbits is near the $x=0$ axis. This is indeed where we would
intuitively place them since the periodic orbits outside corotation
are elongated perpendicular to the bar and since the zero velocity
curves also reach their maximum distance from the center at
$x=0$. Also note that, although the Lyapunov orbit is
unstable, the orbits in the manifold stay in its
close vicinity for at least three bar rotations. By four bar rotation
periods arm stubs are formed, while at five the arms have a winding
of about $3\pi/4$. By six bar rotation periods the manifold has reached
the opposite end of the bar. In other words, the growth is by no means
linear, since growth is slow in the beginning and faster as time
advances. Indeed, it takes more than three orbital periods to leave
the vicinity of the Lyapunov orbit, the first half of the winding is
developed in about $4.5T$ and the full winding in barely at
$6T$. After $6T$, the orbits on the manifold are in the vicinity of
the Lyapunov orbit. These orbits do not leave this region in another
new direction, but they follow the direction given by the
invariant manifolds already formed. 

\begin{figure}
\begin{center}
\includegraphics[scale=0.45,angle=-90]{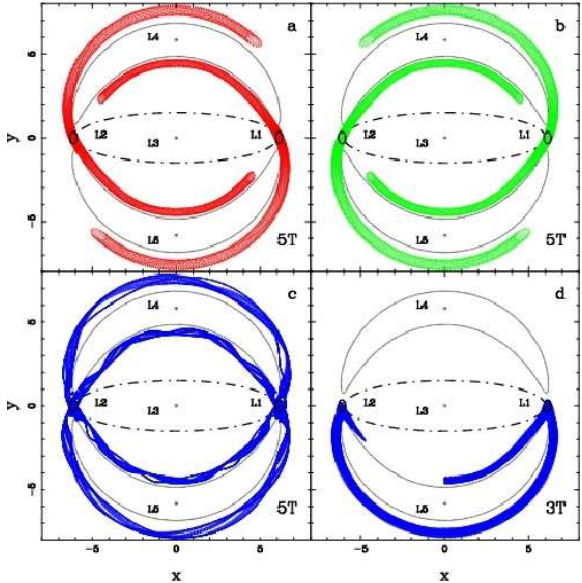}
\end{center}
\caption{Invariant manifolds and perturbations in the plane barred
potential described in section 5 and for $E_J=-0.1977$. {\bf a)} Invariant
unstable manifolds coming from Lyapunov orbits
  around the Lagrangian points $L_1$ and $L_2$. {\bf b)} Invariant
stable manifolds coming from the same
Lyapunov orbits. {\bf c)} Orbits starting from initial conditions near the
  Lyapunov orbit. {\bf d)} Orbits starting from initial conditions
inside the $W^s_{\gamma,1}$ curve of the interior branch of the stable
invariant manifold.}
\label{fig:manifolds}
\end{figure}

Relevant information for a value of the Jacobi constant
$(E_J=-0.1977)$ close to the one of the $L_1 (L_2)$ equilibrium point,
is given in Fig.~\ref{fig:manifolds}. In all four panels,
the bar is outlined by a black dot-dashed line, while the positions
of the five Lagrangian points are marked with  
a star, and the dark grey lines correspond to the zero velocity
curves of this particular value of $E_J$. This means that any orbit
with this energy starting
outside these curves cannot enter within them. We have also plotted
the two plane unstable Lyapunov periodic orbits around 
the unstable Lagrangian points $L_1$ and $L_2$ (black 
solid lines). In Fig. \ref{fig:manifolds}a we show the unstable
invariant manifolds for this value of $E_J$. 
Each of these manifolds is composed of two branches, an interior branch,
lying in the interior region, and
an exterior branch, lying in the exterior region. Each of these
branches can be thought of as an ensemble of orbits moving away from
the Lyapunov orbit. Both branches lie near the zero velocity
curves and, as shown in Fig.~\ref{fig:times}, their length increases with time until they reach the
opposite side of the bar from which they emanated. The interior branch,
when complete, outlines 
well the loci of the inner rings in barred galaxies.
The exterior branch, when complete, has a 
shape similar to that of the $R_1$ outer rings. From a
dynamical point 
of view, in the phase space these branches are seen like tubes that drive the 
dynamics. In Fig. \ref{fig:manifolds}b we plot the stable
invariant manifolds, for the same value of
$E_J$. Again there are two branches, an interior and an exterior one. Note
also that the space loci of the stable and unstable manifolds is almost
identical. There is, however, an important difference in that, for the
stable manifold, the orbits filling these tubes will
go {\it towards} the Lyapunov orbits, while for the unstable manifold
they will go away from it.
In Fig. \ref{fig:manifolds}c
we show four sets of orbits starting from the vicinity of the Lyapunov orbits and
following the four tubes that constitute the two branches of the unstable
manifolds. This figure illustrates how the invariant manifolds drive
the dynamics, since we can see that initial conditions in the vicinity of the
Lyapunov orbit will follow, due to its unstable character, a trajectory close to the unstable
invariant manifolds.  In Fig. \ref{fig:manifolds}d, we represent the
trajectories with initial conditions inside the $W^s_{\gamma,1}$ curve
of the interior branch of the stable invariant manifold, as explained
in the previous section. This set of orbits follows the stable
branch they emanate from, approaching the
Lyapunov orbit, and it leaves the bar region following the exterior
branch of the unstable invariant manifold.

\section{Discussion}

\begin{figure*}
\includegraphics[scale=0.30,angle=-90]{radprf.ps}
\includegraphics[scale=0.30,angle=-90]{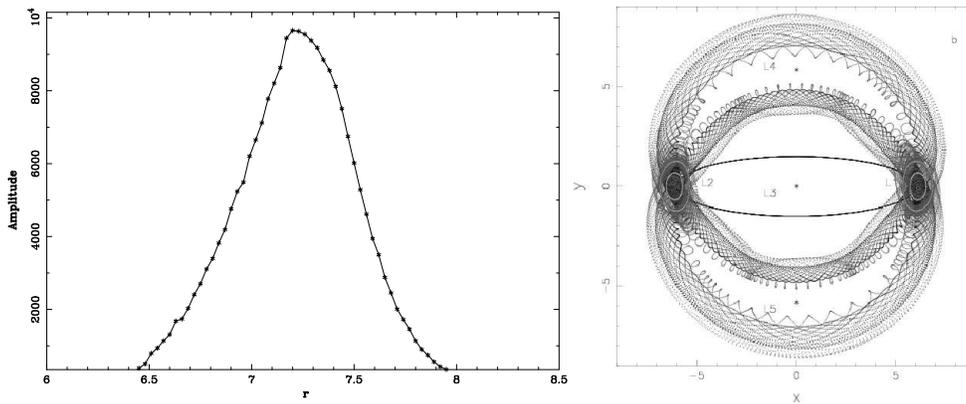}
\caption{{\bf a)} Density profile on a cut across the ring. {\bf b)}
Two unstable invariant manifolds for different values of $E_J$. Note
how similar the regions they delineate are.}
\label{fig:profile}
\end{figure*}

The time during which the orbit stays around the Lyapunov
orbit before outlining the outer or the inner ring depends on the Lyapunov
exponent (Lyapunov \cite{lya49}) of the Lyapunov orbit, and is found to be an increasing
function of $E_J$. Thus, orbits initially near the
Lyapunov orbit with lower values of $E_J$ stay less around the unstable
point. This time increases gradually with the value of $E_J$, until the
value at which the Lyapunov orbit becomes stable
after which all orbits starting in the vicinity of the Lyapunov orbit stay
around it. Thus the orbits outlining the rings are mainly low energy
orbits. Repeating the computations in section 5 for different values of the
Jacobi constant, $E_J=-0.1973$ and $E_J=-0.1960$, we find that the
locus of the invariant manifolds (and
therefore of the orbits associated to it) is roughly independent of their
value of $E_J$. This is illustrated in
Fig. \ref{fig:profile}b, 
for these two different values
of $E_J$. As the energy increases, the size of the Lyapunov
orbit also increases, so that the outline of the invariant manifold
becomes thicker. However, as we consider more orbits and more
energy levels, we find that the density of the central part of the
outlined area increases 
considerably, so that in practice the thickness of the
ring will be considerably smaller than that of the higher energy
manifolds. This is illustrated in Fig.~\ref{fig:profile}a, where we
plot the density profile on a cut across the ring. To obtain this
figure we calculated the unstable invariant 
manifolds and the trajectories inside them for all energy levels
at which the Lyapunov orbit is unstable, i.e. all energies for which
the mechanism we propose can be applied. This covers the range of
values from $E_J=-0.19789$, corresponding to the energy of the unstable
equilibrium point, to the value $E_J=-0.1674$, where the Lyapunov family becomes
stable. The contribution of each energy is weighted
by a distribution function which is here simply assumed to be an
exponentially decreasing function of the energy, i.e. of the form 
$\exp{\left(-\frac{\mid E_J \mid}{2 \sigma^2}\right)}$, with $\sigma=30 kms^{-1}$ as a
velocity dispersion characteristic of disc stars in the solar neighbourhood
(Binney \& Merryfield \cite{binm98}). The exact shape of the
distribution function is of little importance, but it has to be a
decreasing function of the energies in the rotating frame of
reference. The profile shown in
Fig. \ref{fig:profile}a is very similar to those found for the old and
intermediate age stellar population in rings and spirals (Schweizer \cite{schwe76}).

These ring structures will corotate with the bar. During bar
evolution, however, the bar pattern speed will decrease with time due
to an angular momentum exchange (Tremaine \& Weinberg \cite{tre84}; Weinberg
\cite{wei85}; Little \& Carlberg \cite{lit91a}, \cite{lit91b};
Hernquist \& Weinberg \cite{her92}; Athanassoula \cite{ath96};
Debattista \& Sellwod \cite{deb00}; Athanassoula \cite{ath03}; O'Neill
\& Dubinsky \cite{one03}; Valenzuela \& Klypin \cite{val03}). This means that 
$L_1$, $L_2$, $L_4$ and $L_5$ will move outwards and thus 
the reservoir of fresh material for the rings would be continuously 
replenished. This may also be linked to the plumes
that surround the ring structures of these galaxies (Buta
\cite{but84}).

\begin{figure}
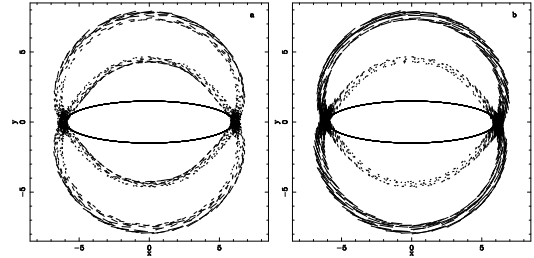

\begin{center}
\includegraphics[scale=0.20,angle=-90]{velfieldrot.ps}
\includegraphics[scale=0.20,angle=-90]{velfieldnon.ps}
\end{center}
\caption{Velocity field along the invariant manifolds. {\bf a)} In the
rotating frame. {\bf b)} In the nonrotating frame.}
\label{fig:velfield}
\end{figure}

The model we propose here forms ring structures which are not 
necessarily associated with the outer Lindblad resonance. In fact, in the
model shown in Fig. \ref{fig:manifolds}, the maximum radius of the inner
branches of the invariants reaches values of about $4.25$ radial units,
while corotation radius is placed at the end of the bar at $6$
radial units. The outer branches extend up to $7.5$ radial units, 
while the Outer Lindblad Resonance of this model is placed at about 
$8.7$ radial units. Note that the ratio of the maximum radius of the 
outer and inner branches give a value of $1.76$ which is in agreement
with the values of this ratio measured for ringed galaxies
(Athanassoula et al. \cite{atha82}; Buta \cite{but95}).

In the rotating frame, the velocities along the invariants reach a
maximum at the point of maximum radius and then decrease toward the
region of the Lyapunov orbits (see Fig. \ref{fig:velfield}a). In the
non-rotating frame the
velocities along the invariants are an order of magnitude higher in
the outer branches than in the inner ones (see
Fig. \ref{fig:velfield}b). A clear prediction from our model is 
that the perturbations on the velocity fields produced by the rings 
should be higher around the outer ring than around the inner ring.
There are little data available on the velocity fiels along the
rings of $R_1$ galaxies. The best data correspond to the galaxy 
NGC 1433 (Buta \cite{but86b}), which is of type $R'_1$.  
For this case, we can tentatively say that there is a general agreement with the
oscillations in the velocities measured in the rotating frame for the 
inner ring. 

Although these rings are not density waves, they 
do not have the shortcoming of material
arms, since they do not wind up with time. A more
appropriate name would be flux rings, since they are outlined by the 
trajectories of particles. The material of such rings would create a
potential well. Other stars and gas in the
galaxy would feel this potential and, while traversing the ring, they
would stay longer at the potential minima, thus adding a density wave
component to the ring. Thus, although we have not made any self
consistent simulations, we can speculate that the flux rings and the density wave 
rings would coincide in the galaxy.  

In this paper we presented a new theory on the origin of $rR_1$
ring structures based on orbital dynamics. We introduced the
invariant manifolds of a periodic orbit, used so far in celestial
mechanics. We explained their role in the transport of
particles from the interior region to the exterior, and viceversa, and
applied it to a realistic barred galaxy model.
Finally, we compared and discussed our results and the characteristics of
the rings obtained with observational data. We can conclude
that the {\sl $rR_1$ ring structure can be interpreted as a bun\-dle 
composed of all the invariant manifolds for all the possible 
energies as well as the orbits driven by them.}

{\bf Acknowledgements}
We thank Albert Bosma for sti\-mulating discussions on properties of
observed rings.
This work is being supported by the spanish MCyT BFM2003-9504 and
catalan 2003XT-00021 grants. JJM also thanks the support of the Agrupaci\'o Astron\`omica de Manresa.

\end{document}